\documentclass[sn-mathphys, Numbered]{sn-jnl}

\usepackage{graphicx}%
\usepackage{multirow}%
\usepackage{amsmath,amssymb,amsfonts}%
\usepackage{amsthm}%
\usepackage{mathrsfs}%
\usepackage[title]{appendix}%
\usepackage{xcolor}%
\usepackage{textcomp}%
\usepackage{manyfoot}%
\usepackage{booktabs}%
\usepackage{algorithm}%
\usepackage{algorithmicx}%
\usepackage{algpseudocode}%
\usepackage{listings}%

\usepackage{amsmath} 

\usepackage{ae}
\usepackage{siunitx}
\usepackage{tikz}
\usetikzlibrary{quantikz}
\usepackage{subcaption}
\usepackage{braket}
\usepackage{amsthm}
\usepackage{amssymb}

\usepackage{comment}

\theoremstyle{thmstyleone}%
\theoremstyle{thmstyletwo}%

\theoremstyle{thmstylethree}%

\begin{document}

\title{
Synergy between noisy quantum computers and scalable classical deep learning
}

\author*[1,2]{\fnm{Simone} \sur{Cantori}}\email{simone.cantori@unicam.it}
\author[1]{\fnm{Andrea} \sur{Mari}}
\author[1,2]{\fnm{David } \sur{Vitali}}
\author[1,2]{\fnm{Sebastiano} \sur{Pilati}}\email{sebastiano.pilati@unicam.it}

\affil[1]{\orgdiv{School of Science and Technology, Physics Division}, \orgname{Universit{\`a}  di Camerino}, \orgaddress{\street{Via Madonna delle Carceri}, \city{Camerino (MC)}, \postcode{62032}, \country{Italy}}}

\affil[2]{\orgdiv{Sezione di Perugia}, \orgname{INFN},  \city{Perugia}, \postcode{06123}, \country{Italy}}

\abstract{
We investigate the potential of combining the computational power of noisy quantum computers and of classical scalable convolutional neural networks (CNNs). The goal is to accurately predict exact expectation values of parameterized quantum circuits representing the Trotter-decomposed dynamics of quantum Ising models. By incorporating (simulated) noisy expectation values alongside circuit structure information, our CNNs effectively capture the underlying relationships between circuit architecture and output behaviour, enabling predictions for circuits with more qubits than those included in the training set. 
Notably, thanks to the quantum information, our CNNs succeed even when supervised learning based only on classical descriptors fails. Furthermore, they outperform a popular error mitigation scheme, namely, zero-noise extrapolation, demonstrating that the synergy between quantum and classical computational tools leads to higher accuracy compared with quantum-only or classical-only approaches. By tuning the noise strength, we explore the crossover from a computationally powerful classical CNN assisted by quantum noisy data, towards rather precise quantum computations, further error-mitigated via classical deep learning.
}


\maketitle
\section{Introduction}
Quantum computers promise to solve computational problems that are intractable on classical machines~\cite{365700, Daley2022}.
However, efforts to exploit the full power of quantum computing are currently limited by hardware errors.
To address this issue, quantum error mitigation techniques have been developed to minimize noise and obtain potentially useful results~\cite{cai2023quantum,ibm2023,Kim2023,PhysRevLett.119.180509,vandenBerg2023,PRXQuantum.2.040330}.
While error mitigation methods reduce noise in expectation values of observables, they may display limited accuracy or suffer from prohibitive sampling overheads \cite{takagi2023universal, quek2022exponentially}.
In this scenario, classical machine learning emerges as a suitable tool for post-processing noisy quantum measurements, achieving accurate expectation values at a potentially lower computational cost~\cite{ibmML,PhysRevResearch.6.013223}. In fact, supervised machine learning has been successfully applied to various challenging computational tasks within quantum many-body physics~\cite{Huang_2022,RevModPhys.91.045002,schutt2020machine,Kulik_2022,PRXQuantum.2.040201} and quantum computing~\cite{Baireuther_2019,Baireuther2018machinelearning,Chamberland_2018,torlai2018neural,encoding,Cantori_2023,ibmML,mohseni2023,cantori2024challenges,melko2024language}.
Moreover, scalable supervised learning models allow generalizing beyond the size of the training quantum systems, potentially reaching system sizes out of reach for direct classical simulations~\cite{C8SC04578J,Saraceni,https://doi.org/10.1002/syst.201900052,10.21468/SciPostPhys.10.3.073,Mohseni_2023}.
One the other hand, classically supervised learning was shown to fail in emulating certain relevant quantum circuits~\cite{mohseni2023}, e.g., circuits featuring random inter-layer variations~\cite{cantori2024challenges}.
%

\begin{figure*}[]
	\centering
	\includegraphics[width=1\textwidth]{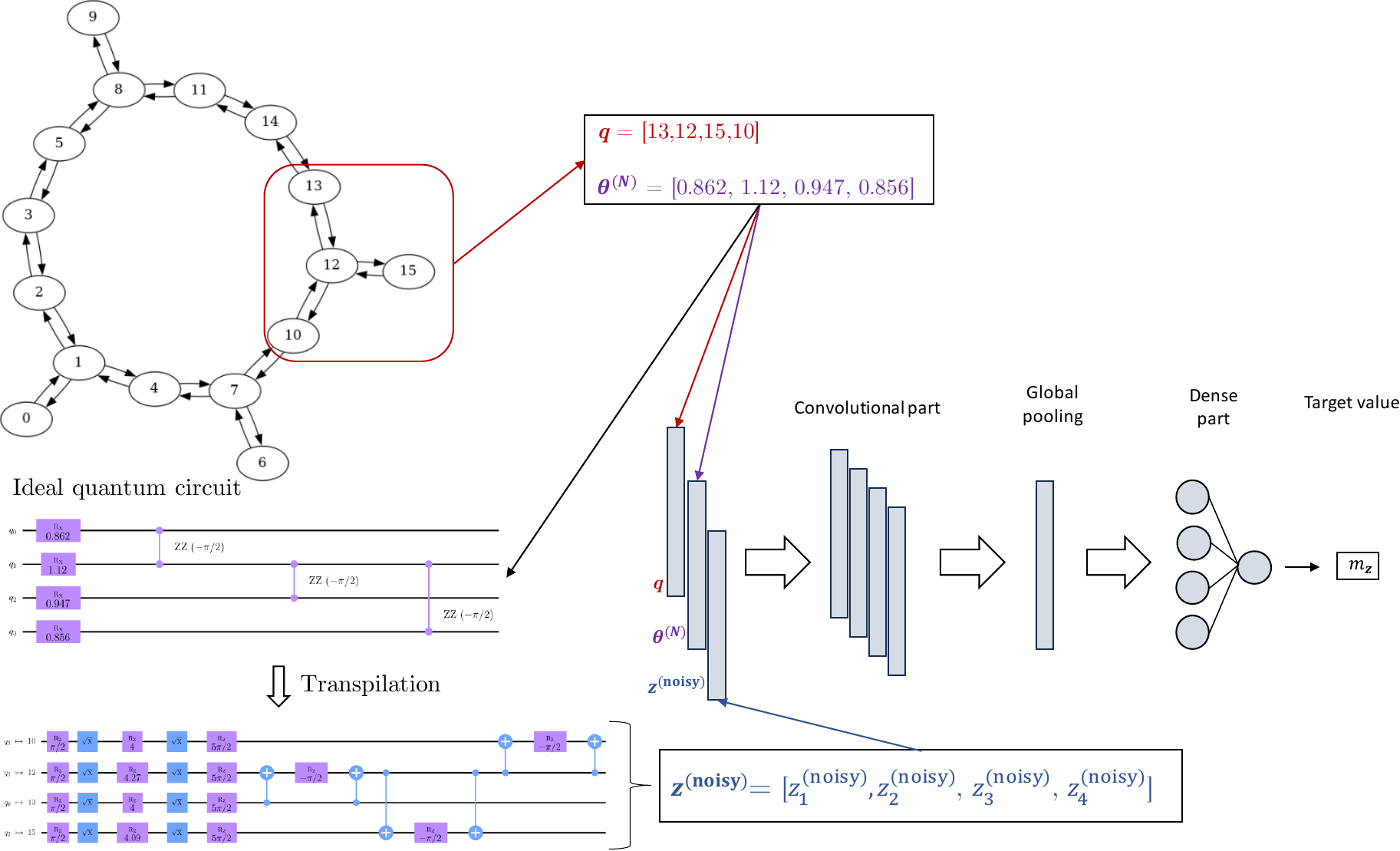}
	\caption{Schematic representation of the synergetic computation combining classical deep learning with output of noisy quantum circuits. In this example, a quantum circuit with $N=4$ qubits and $P=1$ layers is considered. The structure of the IBM Guadalupe chip is shown in the upper part. Sections of $N$ adjacent qubits are randomly selected, and the corresponding indices are denoted with $\boldsymbol{q}$.  
 The single-qubit rotation angles $\boldsymbol{\theta}^{(N)/(P)}=[\theta_1,\theta_2,...,\theta_{N/P}]$ are randomly generated from a uniform distribution in the range $\theta_i \in [0,\frac{\pi}{2}]$. The logical circuits feature single-qubit rotations and CNOT gates. They are transpiled for the quantum chip layout and its basis gates. The noisy expectation values $\boldsymbol{z}^{\mathrm{(noisy)}}$ of the transpiled circuits are the input to a CNN, together with the classical circuit descriptors $\boldsymbol{q}$ and  $\boldsymbol{\theta}^{(N)/(P)}$. This supervised learning model is trained to predict the exact average magnetization per qubit $m_{\boldsymbol{z}} = \frac{1}{N}\sum_{n=1}^{N}\left<\psi_{\textrm{out}}\right| Z_n \left| \psi_{\textrm{out}} \right>$, where $\left|\psi_{\mathrm{out}}\right>$ is the output state of the quantum circuit.
}
	\label{Fig1}
\end{figure*}

In this work, we investigate the computational synergy between noisy quantum computers and classical deep learning. Specifically, our focus is on the task of predicting expectation values of large quantum circuits representing the Trotter-decomposed dynamics of an Ising Hamiltonian~\cite{ibm2023,ibmML,narasimhan2023simulating}. These circuits are simulated taking into account the connectivity of an actual quantum chip and considering a realistic model of hardware errors.
Our approach involves incorporating noisy quantum expectation values alongside information on the circuit architecture, to be used as input features for classical neural networks. 
A schematic representation is shown in Fig.~\ref{Fig1}.
Leveraging scalable network generalization, our method shows remarkable performance in emulating quantum circuits with more qubits than those included in the training set.
In this way, our approach also performs accurate quantum error mitigation, but circumventing the need of explicit target values for large circuits. Thus, it departs from the requirement of error-mitigated expectation values as training data~\cite{ibmML}. 
On the other hand, our investigation improves upon the practice of relying only on circuit-structure information for predicting expectation values~\cite{Zhang_2021,Cantori_2023, mohseni2023, cantori2024challenges}. Notably, this allows us emulating circuits that are otherwise intractable for purely classical supervised learning.
This approach outlines the potential of combining the outputs provided by quantum computers and classical deep learning methods. The synergy between these two strategies promises results that surpass the individual capabilities of each. 

The rest of the article is organized as follows: in Sec.~\ref{Sec2} we describe the quantum circuits we address and the structure of the quantum chip on which they can be implemented. We also introduce the error model used to simulate the noisy expectation values, as well as the technique we implement to tune the noise level. The CNNs and the training protocol are described in the final part of the section. 
The scalability of the CNNs on larger quantum circuits is analysed in Sec.~\ref{Sec3}. Here, we compare the accuracy of the predictions for different quantum circuit configurations, different numbers of qubits, and different levels of noise. Notably, comparison is made also against a prominent error-mitigation technique, namely zero-noise extrapolation (ZNE)~\cite{PhysRevLett.119.180509,kandala2019error, li2017efficient}. In Sec.~\ref{Sec4} we report our conclusions. 
Further details on how we tune the noise model and how we implement ZNE are available in Appendix~\ref{appendix_noise} and Appendix~\ref{appendixZNE}, respectively.

\section{Methods}\label{Sec2}
\subsection{Quantum circuits and qubit arrangement}
\begin{figure*}[]
	\centering
	\includegraphics[width=\textwidth]{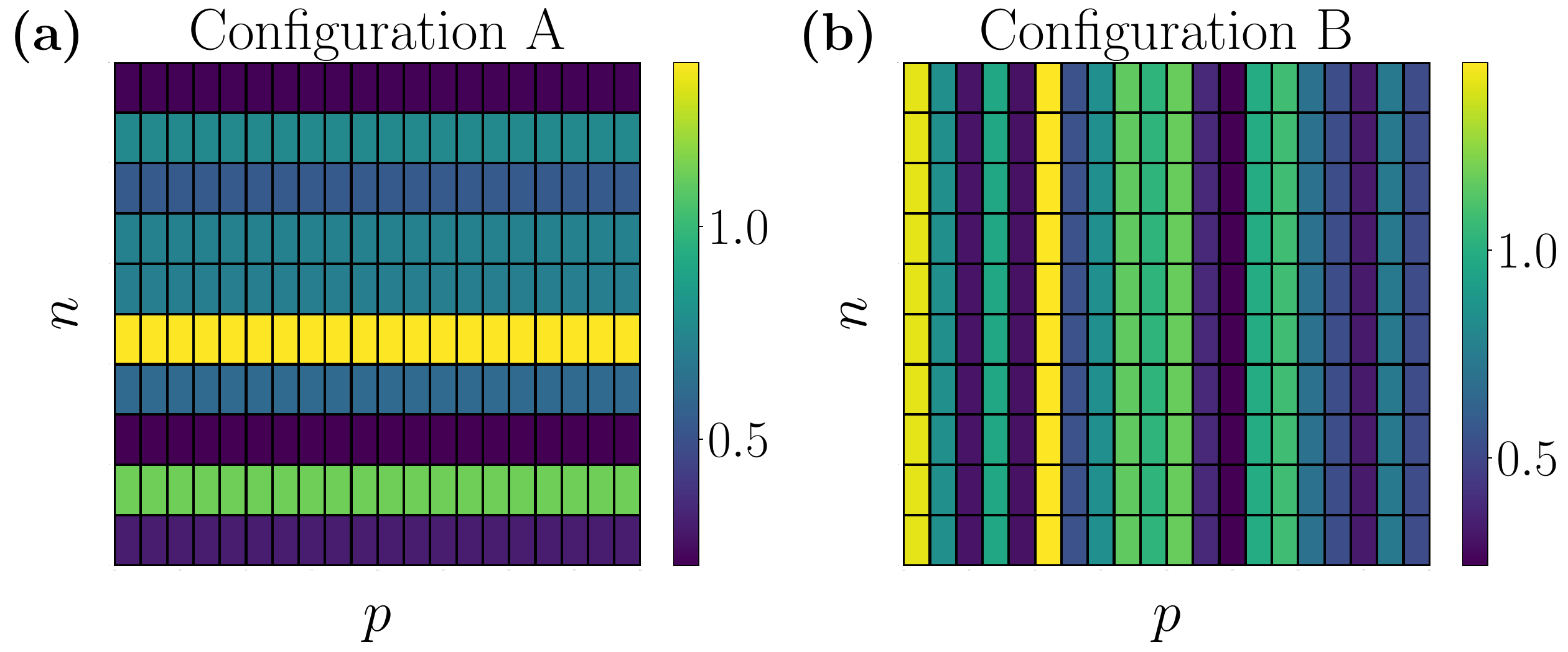}
	\caption{Color-scale representation of the rotation angles describing the $R_X$ gates ($\boldsymbol{\theta}^{(N)}$ and $\boldsymbol{\theta}^{(P)}$) as a function of the qubit index $n=1,\dots,N$ and the layer index $p=1,\dots,P$. The angles are sampled from a uniform distribution within the interval $[0, \frac{\pi}{2}]$. Panel (a) shows a quantum circuit in configuration A, where angles are different for different qubits. Panel (b) shows a quantum circuit in configuration B, where different layers of gates feature distinct angles.
}
	\label{configurations}
\end{figure*}

We consider quantum circuits composed of $N$ qubits and $P$ layers of gates. In each
layer, a parameterized single-qubit gate $R_X$ is applied to each qubit, and two-qubit gates $R_{ZZ}$ are applied to chosen qubit pairs.
The matrix representations of these gates are:
\begin{align}
    R_X(\theta) = 
    \begin{bmatrix}
        \cos(\frac{\theta}{2}) & -i\sin(\frac{\theta}{2}) \\
        -i\sin(\frac{\theta}{2}) & \cos(\frac{\theta}{2})
    \end{bmatrix},
    &&
    R_{ZZ}(\phi) = 
    \begin{bmatrix}
        e^{-i\frac{\phi}{2}} & 0 & 0 & 0 \\
        0 & e^{i\frac{\phi}{2}} & 0 & 0 \\
        0 & 0 & e^{i\frac{\phi}{2}} & 0 \\
        0 & 0 & 0 & e^{-i\frac{\phi}{2}}
    \end{bmatrix}.
\end{align}
This type of quantum circuit can be used to simulate the time dynamics of a many-body quantum system described by the transverse-field Ising Hamiltonian, which is defined as:
\begin{equation}
    H(t)=H_{ZZ} + H_{X} = -J\sum_{\langle i,j \rangle}Z_iZ_j + \sum_i h_i(t)X_i \, ,
\end{equation}
where $X_i$ and $Z_i$ are Pauli operators, $J$ is the coupling between nearest-neighbour spins on the chosen graph, and $h_i(t)$ is the time-dependent transverse field acting on qubit $i$. Indeed, from the  first-order Trotter decomposition of the time-evolution operator, we get
\begin{align}
    \mathrm{e}^{-\mathrm{i}H_{ZZ}\delta t} &= \prod_{\langle i,j \rangle} \mathrm{e}^{\mathrm{i}J\delta t Z_i Z_j} = \prod_{\langle i,j \rangle} \mathrm{R}_{Z_iZ_j}(-2J\delta t)\\
    \mathrm{e}^{-\mathrm{i}H_{X}\delta t} &= \prod_{i} \mathrm{e}^{-\mathrm{i}h(t)\delta t X_i} = \prod_{i} \mathrm{R}_{X_i}(2h(t)\delta t) \, ,
\end{align}
where the total evolution time $T$ is discretized into $\frac{T}{\delta t}$ Trotter steps, $-2J\delta t = \phi$, and $2h(t)\delta t = \theta$.
We set $\phi=-\frac{\pi}{2}$, following the approach of Ref.~\cite{ibm2023}, despite employing a different circuit transpilation method.
The angles $\theta$ for the $R_X$ gates are randomly selected from a uniform distribution within the interval $[0, \frac{\pi}{2}]$. 
As shown in Fig.~\ref{configurations}, we consider two distinct circuit configurations: A and B. In configuration A, the angles are randomly assigned to each qubit, but the same angle set is used across the $P$ layers of gates. Instead, in configuration B the single-qubit gates feature different angles for different layers, but the angles are consistent across qubits within a specific layer.

The qubit pairs connected by the $R_{ZZ}$ gates consist exclusively of the nearest neighbours on the graph of the IBM Guadalupe chip. This is illustrated in Fig.~\ref{Fig1}. 
Specifically, different portions of the chip are considered in different random realizations of the parametrized circuit. We consider all the possible connections of the quantum chip except the one between the physical qubit 4 and the physical qubit 1 (open boundary conditions).
Therefore, each realization is uniquely determined by two arrays. The first, indicated with $\boldsymbol{q}$, includes the indices labelling the physical qubits selected in the considered circuit realization (see Fig.~\ref{Fig1}). This information is important for identifying the connections among qubits. The second array is the set of angles $\boldsymbol{\theta}^{(N)} = \{\theta_1, \theta_2, ..., \theta_N\}$ for configuration A, or $\boldsymbol{\theta}^{(P)} = \{\theta_1, \theta_2, ..., \theta_P\}$ for configuration B.
To accurately model the noise characteristics of the IBM Guadalupe chip, we need to transpile our ideal quantum circuit into a form that can be executed on the quantum device, using the available gate set.
This process is performed by Qiskit and is visualized in Fig.~\ref{Fig1}.
While the arrays $\boldsymbol{\theta}$ and $\boldsymbol{q}$ uniquely identify each circuit realization and, hence, are suitable to perform purely classical supervised learning, we augment the circuit description with the set of noisy expectation values that would be produced by noisy quantum circuits, as discussed hereafter.

\subsection{Noisy expectation values}\label{subsec_noise}
The target value our CNNs shall predict is the average magnetization per qubit:
\begin{equation}
    m_{\boldsymbol{z}} = \frac{1}{N}\sum_{n=1}^{N} z_n \equiv  \frac{1}{N}\sum_{n=1}^{N}\left<\psi_{\textrm{out}}\right| Z_n \left| \psi_{\textrm{out}} \right> \, ;
\end{equation}
%
$\left|\psi_{\textrm{out}}\right>$ is the output state of the quantum circuit after the application of $P$ layers of gates on the input state $\left|\psi_{\textrm{in}}\right>=\left|0\right>^{\otimes N}$.
For each circuit, the target value is exactly determined via state-vector simulations. 
We also numerically emulate the  execution of a noisy quantum computer. For this, we adopt the noise model encoded in the virtual backend \textit{FakeGuadalupe} available in the Qiskit library~\cite{qiskit}. This model replicates the noise characteristics of the original IBM Guadalupe quantum chip. The noisy quantities corresponding to the exact single-qubit expectation values $z_n$, for $n=1,\dots,N$, will be collectively denoted  as $\boldsymbol{z}^{\mathrm{(noisy)}}=\{z_1^{\mathrm{(noisy)}}, z_2^{\mathrm{(noisy)}}, ..., z_N^{\mathrm{(noisy)}}\}$.
 These noisy expectation values might help the networks predicting the corresponding ground-truth results. Hence, we provide them as a further input to the CNNs, in addition to the classical circuit descriptors $\boldsymbol{\theta}$ and $\boldsymbol{q}$. 
 This combination of (here, simulated) quantum data and classical circuit-features allows overcoming previous approaches that either used classical descriptors only, or error-mitigated same-size circuit outputs, without exploiting scalable classical networks.
 Clearly, with our approach we aim to obtain predictions that at least outperform the accuracy of the trivial estimation:
\begin{equation}
m_{\boldsymbol{z}^{\mathrm{(noisy)}}}=\frac{1}{N}\sum_{n=1}^{N} z_n^{\mathrm{(noisy)}}. 
\end{equation}

In the following, it will be useful to tune the amount of noise in the circuit outputs. Specifically, we choose to focus on the errors associated with the CNOT gates, which are dominant compared to other errors, e.g. those associated with the single-qubit rotations or with readout operations. Our procedure to tune the noise level is described in Appendix~\ref{appendix_noise}.
In short, we introduce the parameter $p_{\mathrm{noise}}$, with $1\ge p_{\mathrm{noise}} \ge 0$, which determines the noise strength associated to the CNOT gates. The value $p_{\mathrm{noise}}=1$ corresponds to the standard noise model of the quantum chip, while $p_{\mathrm{noise}}=0$ indicates the total cancellation of the noise related to the CNOT gates. 
Notice that, while somewhat less effective, other errors such as the ones on the other gates and readout errors are still allowed.


\subsection{Convolutional neural networks}
As discussed in the previous sections, we train deep CNNs to predict expectation values $m_{\boldsymbol{z}}$ of different quantum circuits. For quantum circuits in configuration A, the network input is one dimensional and it features three channels, resulting in the input shape $(N,3)$. The first channel includes the qubit indices $\boldsymbol{q}$\footnote{The actual descriptors are normalized as $\boldsymbol{q}^{\prime}=\boldsymbol{q}/10$, so that values in different channels are of the same order of magnitude. With more qubits, a higher normalization factor might be appropriate.}, the second one includes the angles $\boldsymbol{\theta}^{(N)}$, while the third channel includes the noisy expectation values $\boldsymbol{z}^{\mathrm{(noisy)}}$. These three channels allow the CNNs combining classical circuit descriptors with noisy quantum data. 
For circuits in configuration B, we implement a two-dimensional CNN with input shape $(N,P,3)$. 
To fit this shape, the length-$P$ array $\boldsymbol{\theta}^{(P)}$ is repeated $N$ times. Both $\boldsymbol{q}$ and $\boldsymbol{z}^{\mathrm{(noisy)}}$ are repeated $P$ times for the same reason.
We compare the performance of these CNNs with analogous networks that process only the classical circuit descriptors, namely $\boldsymbol{\theta}^{(N)/(P)}$ (for configuration A/B) and $\boldsymbol{q}$.
In these cases, the networks have two input channels. To distinguish the above models, we respectively indicate the network with hybrid classical-quantum inputs  with CNN($\boldsymbol{\theta}^{(N)/(P)}$, $\boldsymbol{q}$, $\boldsymbol{z}^{\mathrm{(noisy)}}$), and the one with only classical descriptors with CNN($\boldsymbol{\theta}^{(N)/(P)}$, $\boldsymbol{q }$).

Our final goal is to predict expectation values of quantum circuits larger than those included in the training set. To adapt the network to the different circuit sizes, a scalable architecture is crucial. Conventional CNNs featuring convolutional filters followed by dense layers are not entirely scalable. Indeed, while convolutional layers can handle variable-sized inputs, dense layers necessitate a fixed input size. To overcome this constraint, we incorporate a global pooling operation after the last convolutional layer, emulating the strategy employed in Refs.~\cite{Saraceni,Cantori_2023}. This enhancement transforms the architecture into a fully scalable framework.
Moreover, consistently training the neural network on a fixed set of physical qubits poses a challenge. Indeed, when tested on larger circuits, the network would encounter configurations involving connections among qubits that were not part of its training data, making scalability impractical. To address this limitation, the CNN is trained on circuits implemented on randomly-selected consecutive portions of the Guadalupe chip, as illustrated in Fig.~\ref{Fig1}. In other words, the CNN is trained with varying combinations of $\boldsymbol{q}$.

The training of the CNN is performed by minimizing the mean squared error loss-function:
\begin{equation}\label{mse}
	\mathcal{L} = \frac{1}{K_\mathrm{train}}\sum_{k=1}^{K_\mathrm{train}}
 \left(y_k - \tilde{y}_k \right)^2 \, ,
\end{equation}
where $K_{\mathrm{train}}$ is the number of instances included in the training set, $y_k=m_{\boldsymbol{z},k}$ is the target value, and $\tilde{y}_k$ is the corresponding predicted value.
The network parameters are optimized via a widely used form of stochastic gradient descent, namely, the ADAM algorithm~\cite{adam}.

To assess the prediction accuracy, we evaluate the coefficient of determination
\begin{equation}\label{r2}
R^2=
1-\frac{
\sum_{k=1}^{K_\mathrm{test}} \left(y_k - \tilde{y}_k \right)^2}{\sum_{k=1}^{K_\mathrm{test}} \left(y_k - \bar{y} \right)^2} ,
\end{equation}
where $\bar{y}$ is the average of the target values and $K_{\mathrm{test}}$ is the number of instances in the test set.
In the following, it will be useful to estimate the correlation between noisy expectation values and the exact ones. For this, we determine the Pearson correlation coefficient:
\begin{equation}\label{pearson}
    \rho = \frac{ \sum_i (m_{\boldsymbol{z}}^{(i)}-\overline{m_{\boldsymbol{z}}})(m_{\boldsymbol{z}^{\mathrm{(noisy)}}}^{(i)}-\overline{m_{\boldsymbol{z}^{\mathrm{(noisy)}}}}) }{\sqrt{\sum_i (m_{\boldsymbol{z}}^{(i)}-\overline{m_{\boldsymbol{z}}})^2\sum_i(m_{\boldsymbol{z}^{\mathrm{(noisy)}}}^{(i)}-\overline{m_{\boldsymbol{z}^{\mathrm{(noisy)}}}})^2}} \, .
\end{equation}
In Eq.~\eqref{pearson}, $\overline{m_{\boldsymbol{z}}}$ and $\overline{m_{\boldsymbol{z}^{\mathrm{(noisy)}}}}$ represent the average of $m_{\boldsymbol{z}}$ and $m_{\boldsymbol{z}^{\mathrm{(noisy)}}}$ across the selected sample of quantum circuits.

\section{Results and Discussion}\label{Sec3}
\subsection{Quantum circuits in configuration A}
\begin{figure*}
	\centering
	\includegraphics[width=\textwidth]{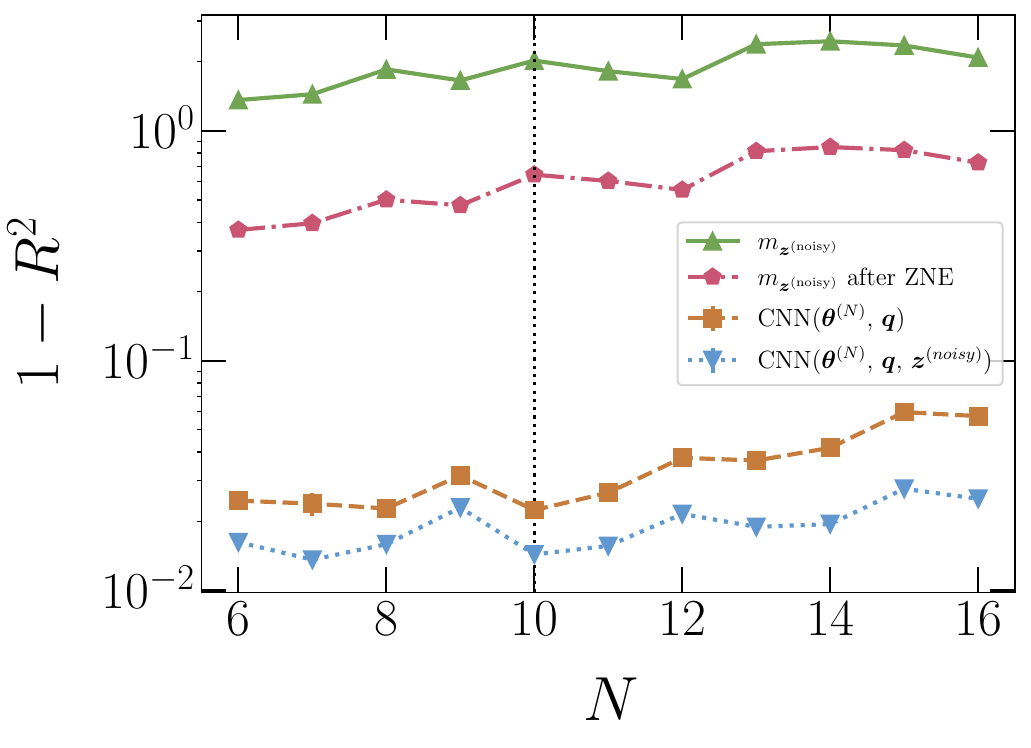}
	\caption{Prediction error $1-R^2$ as a function of the number of qubits $N$ of the quantum circuits in the test set. We compare the network processing only classical descriptors, namely, CNN($\boldsymbol{\theta}^{(N)}$, $\boldsymbol{q}$), the one processing also (simulated) noisy quantum outputs CNN($\boldsymbol{\theta}^{(N)}$, $\boldsymbol{q}$, $\boldsymbol{z}^{\mathrm{(noisy)}}$), as well as the noisy expectation values $m_{\boldsymbol{z}^{\mathrm{(noisy)}}}$ and the corresponding results after zero-noise extrapolation (ZNE). The CNNs are trained on $K_{\mathrm{train}}\simeq 6\times 10^5$ quantum circuits with $N\le 10$ qubits (see  vertical dotted line). The depth of the quantum circuits is $P=20$. 
 }
	\label{varN_extrap}
\end{figure*}

\begin{figure*}
	\centering
	\includegraphics[width=\textwidth]{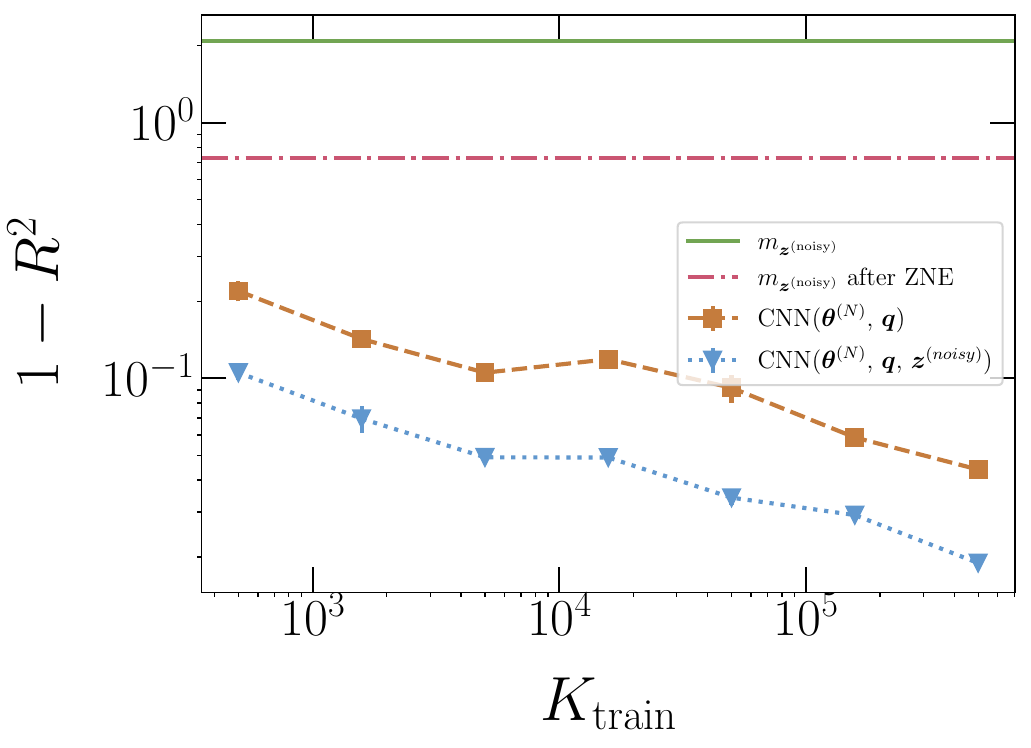}
	\caption{Prediction error $1-R^2$ as a function of the number of instances in the training set $K_{\mathrm{train}}$. The CNNs are trained on quantum circuits in configuration A, $N\le 10$ qubits and $P=20$ layers of gates. The test is performed on quantum circuits with $N=16$ qubits. The different datasets are defined as in Fig.~\ref{varN_extrap}.
 }
	\label{Ktrain_varN}
\end{figure*}

The first test we discuss is on quantum circuits of depth $P=20$ in configuration A. In this scenario, the CNN is trained using quantum circuits with $N\in \{6,...,10 \}$ qubits. Next, the network is tested on quantum circuits featuring up to $N=16$ qubits.
Fig.~\ref{varN_extrap} shows the prediction accuracy as a function of the number of qubits in the test circuits. Here and for the remaining results, the error bars represent the estimated standard deviation of the average over three repetitions of the training process.
We observe that the network which processes only classical circuit descriptors, namely, CNN($\boldsymbol{\theta}^{(N)}$, $\boldsymbol{q}$), achieves satisfactory accuracies. Analogous findings have been previously reported in Ref.~\cite{cantori2024challenges} for a similar circuit structure. However, the hybrid network CNN($\boldsymbol{\theta}^{(N)/(P)}$, $\boldsymbol{q}$, $\boldsymbol{z}^{\mathrm{(noisy)}}$), which processes also the noisy quantum expectation values $\boldsymbol{z}^{\mathrm{(noisy)}}$, consistently reaches superior performances. 
Importantly, we observe that both CNNs outperform the output of the simulated quantum computer, even when the noise is mitigated through ZNE.
In Fig.~\ref{Ktrain_varN}, we show the performance of the CNNs, tested on the qubit number $N=16$, as a function of the number of instances in the training set $K_{\mathrm{train}}$. Notably, the accuracy of the CNNs are better than the ones obtained with the simulated quantum chip even for training sets as small as $K_{\mathrm{train}}\simeq 500$. 

It is worth emphasizing that, in the approach envisioned here, there is no sampling overhead during the prediction phase. In other words, once the network has been trained, for each testing circuit we use the same number of measurements (and even the same noisy results) that are required for the trivial direct estimation of the average  magnetization. Moreover, apart from the negligible classical computing cost of computing the output of the CNN, no classical simulation of test circuits is required. Furthermore, during the training phase, only small-scale circuits must be classically simulated, meaning that large-scale simulations at the size of the test circuits are never required.
%
%

\subsection{Quantum circuits in configuration B}
\begin{figure*}
	\centering
	\includegraphics[width=\textwidth]{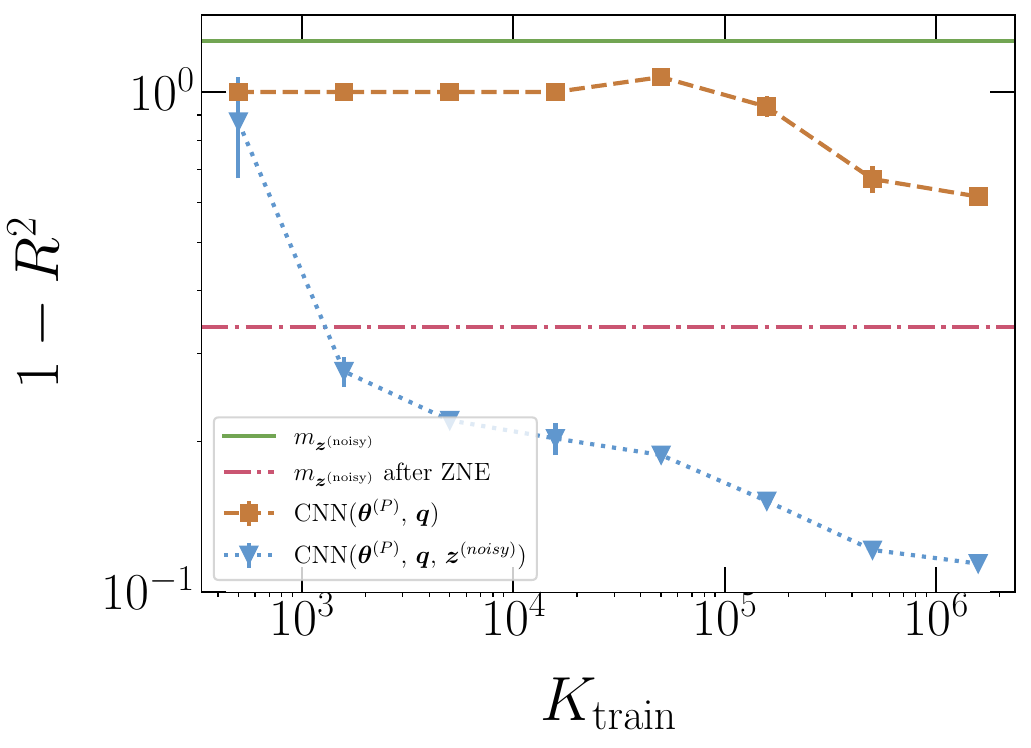}
	\caption{Prediction error $1-R^2$ as a function of the number of instances in the training set $K_{\mathrm{train}}$. The CNNs are trained on quantum circuits in configuration B, featuring $N\le 10$ qubits and $P=20$ layers of gates. The test is performed on quantum circuits with $N=16$ qubits. The different datasets are defined as in Fig.~\ref{varN_extrap}.
 }
	\label{Ktrain_varP_mit100}
\end{figure*}

\begin{figure*}
	\centering
	\includegraphics[width=\textwidth]{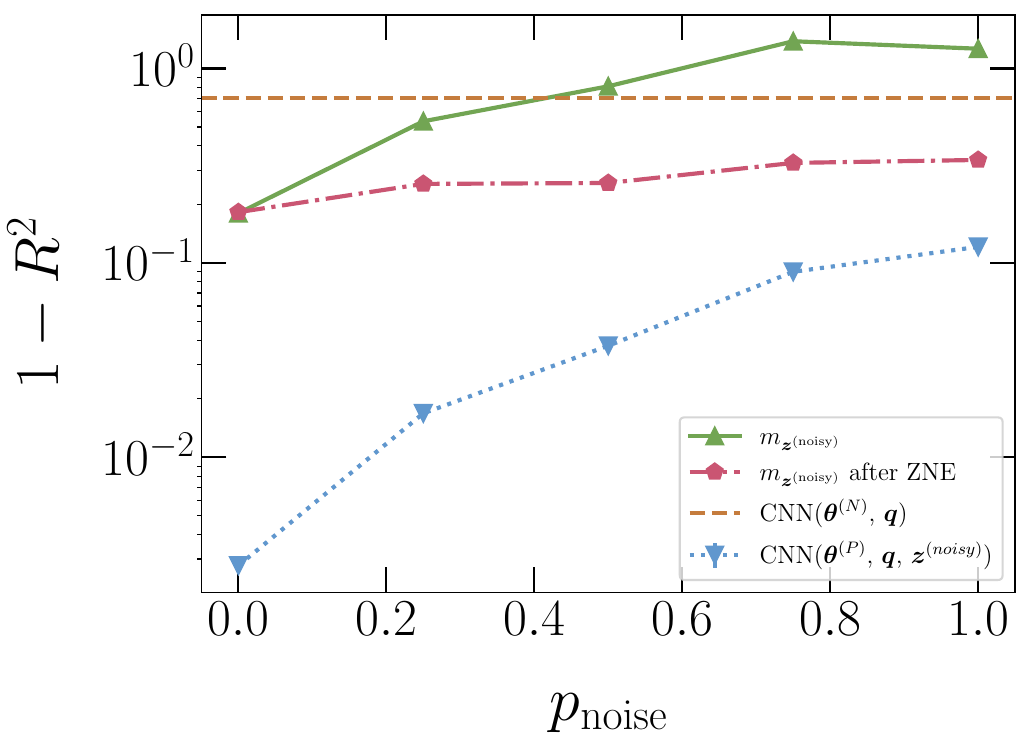}
	\caption{Prediction error $1-R^2$ as a function of the level of noise $p_{\mathrm{noise}}$ associated with the CNOT gates. The CNNs are trained on $K_{\mathrm{train}}\simeq6\times 10^5$ quantum circuits in configuration B, featuring $N\le 10$ qubits and $P=20$ layers of gates. The test is performed on quantum circuits with $N=16$ qubits. 
 The different datasets are defined as in Fig.~\ref{varN_extrap}.
 }
	\label{pnoise}
\end{figure*}

\begin{figure*}
	\centering
	\includegraphics[width=\textwidth]{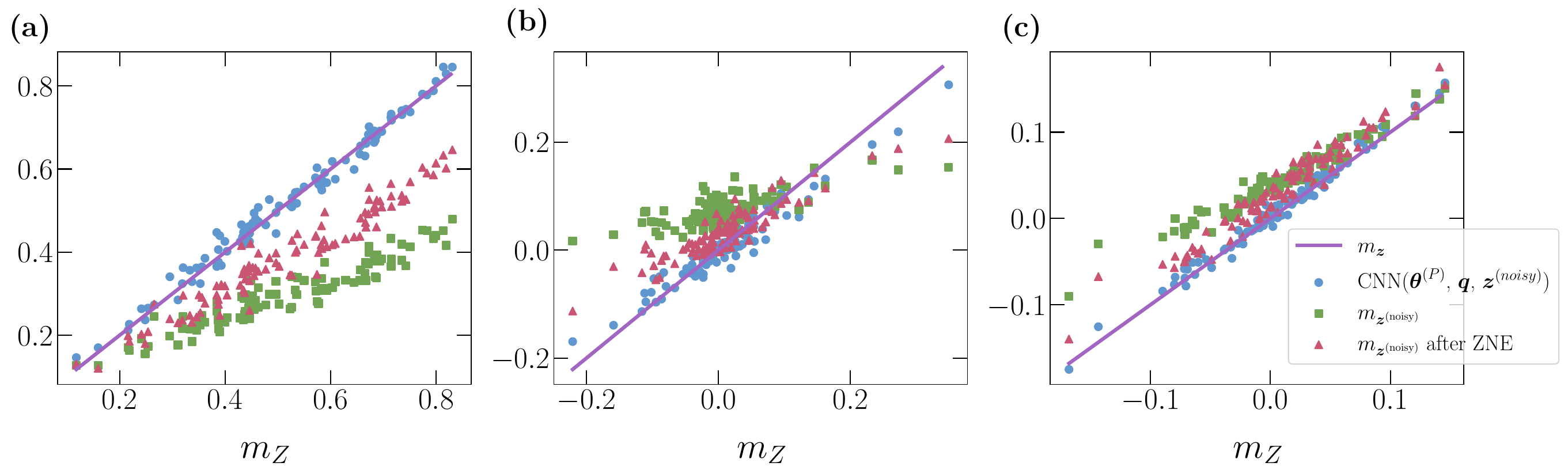}
	\caption{Scatter plot of predictions versus ground-truth expectation values $m_{\boldsymbol{z}}$ for quantum circuits with $N=16$ qubits. The (brown) solid line represents the bisector $\tilde{y}=m_{\boldsymbol{z}}$. The (blue) circles represent the predictions of CNN($\boldsymbol{\theta}^{(N)/(P)}$, $\boldsymbol{q}$, $\boldsymbol{z}^{\mathrm{(noisy)}}$), trained on $K_{\mathrm{train}}\simeq6\times 10^5$ quantum circuits. The (green) squares represent noisy expectation values $m_{\boldsymbol{z}^{\mathrm{(noisy)}}}$. The (red) triangles are the noisy expectation values mitigated via ZNE. (a) The quantum circuits are in configuration A. They feature $P=20$ layers of gates and the CNN is trained on circuits with $N\le 10$ qubits. (b) The quantum circuits are in configuration B with $P=20$. The CNN is trained on circuits with $N\le 10$.
    (c) Same as in (b) but with $p_{\mathrm{noise}}=0.25$.
 }
	\label{zne}
\end{figure*}

It was recently shown that classical neural networks trained via supervised learning fail to emulate quantum circuits featuring rapid random inter-layer angle fluctuations~\cite{cantori2024challenges}.
This failure is replicated here for quantum circuits in configuration B, as shown in Fig.~\ref{Ktrain_varP_mit100}. Indeed, the network CNN($\boldsymbol{\theta}$, $\boldsymbol{q}$), which processes only classical inputs, fails to reach reasonable accuracies $R^2 \simeq 1$, even with as many as $K_{\mathrm{train}}\simeq 10^6$ training circuits (training sizes $N=6,\dots,10$, testing size $N=16$). 
In this test, the advantage of including noisy expectation values $\boldsymbol{z}^{\mathrm{(noisy)}}$ is extreme. Indeed, we find that the network with hybrid inputs, namely, CNN($\boldsymbol{\theta}^{(P)}$, $\boldsymbol{z}^{\mathrm{(noisy)}}$, $\boldsymbol{q}$), produces results with acceptable accuracies. In fact, it outperforms the accuracy of ZNE already with $K_{\mathrm{train}} \gtrsim 10^3$ training circuits.
Still, the accuracy is inferior to the one obtained for configuration A. 
This might be attributed to a lower correlation between the noisy expectation values $m_{\boldsymbol{z}^{\mathrm{(noisy)}}}$ and the ground-truth values $m_{\boldsymbol{z}}$. In fact, the corresponding Pearson correlation coefficient for quantum circuits in configuration A with, e.g., $N=16$ and $P=20$, is $\rho=0.945$, while for quantum circuits of the same size in configuration B it is only $\rho=0.664$. 
Hence, it is natural to ask if and how much the predictions of the CNN which processes also $\boldsymbol{z}^{\mathrm{(noisy)}}$, beyond the classical descriptors, improve when the quantum hardware is less affected by noise.
We analyse this effect by reducing the amount of errors associated to the CNOT gates, as discussed in Sec.~\ref{Sec2} (see Appendix~\ref{appendix_noise} for further details). 
The prediction accuracy is shown in Fig.~\ref{pnoise}, as a function of the noise level $p_{\mathrm{noise}}$. We iterate that the errors beyond those associated to the CNOT gates are not tuned compared to the default \textit{FakeGuadalupe} model. Interestingly, we find that even small improvements in the noisy quantum data lead to a substantial boost in the accuracy of CNN($\boldsymbol{\theta}$, $\boldsymbol{q}$, $\boldsymbol{z}^{\mathrm{(noisy)}}$). Chiefly, this model   systematically outperforms the network with only classical inputs CNN($\boldsymbol{\theta}$, $\boldsymbol{q}$), as well as the ones corresponding to the noisy expectation values $\boldsymbol{z}^{\mathrm{(noisy)}}$, even when these are corrected via ZNE.
It is worth mentioning that at $p_{\mathrm{noise}}=0$ ZNE does not affect the result. This is because certain types of noise, like readout errors, cannot be addressed using this error mitigation technique.
%
To further visualize the comparison between the CNN predictions and the noisy outputs of the simulated quantum chip, in Fig.~\ref{zne} we show scatter plots of the average magnetizations per qubit for a representative testing circuit size. 
For configuration A, one notices an appreciable correlation between noisy expectation values and ground-truth results. The correlation is less pronounced for configuration B. Furthermore, in the latter case the noisy expectation values are rather concentrated, and this contributes to the difficulty of the discriminative learning task. 
%
%

The above findings underscore the promising synergy between classical deep learning and quantum circuit outputs. Noisy expectation values offer valuable insights to the neural networks, enabling them to predict expectation values significantly more accurately, even in setups where supervised learning with only classical descriptors drastically fails. Meanwhile, employing CNNs to mitigate noisy expectation value errors yields superior accuracies compared to those achieved with simulated noisy quantum computers, even when using a prominent error mitigation technique such as ZNE.

It is useful to discuss our approach vis-\`a-vis the machine-learning technique for quantum error mitigation discussed in Ref.~\cite{ibmML}. The significant distinction lies in the training method and in the scope of the network. In Ref.~\cite{ibmML}, the size of the training circuits is equal to the size of the test circuits and zero-noise extrapolated expectation values obtained from a quantum computer are used as training targets. In fact, the main goal of Ref.~\cite{ibmML} is not outperforming the accuracy of ZNE, but rather reproducing equivalent results with a reduced sampling overhead. 
%
In contrast, our scalable architecture eliminates the need to train the neural network directly on large quantum circuits and, consequently, it can be trained  with exact target values associated to small-scale circuits. Due to the different training method, our model can be used as a way of reducing the sampling overhead but also as a way of improving the estimation accuracy compared to standard error mitigation. Indeed, in our numerical simulations, we observe a better accuracy  compared with ZNE, despite paying the same sampling cost of direct estimation.


\section{Conclusions}\label{Sec4}
In this work, we spotlighted the effectiveness of combining scalable classical neural networks with noisy quantum computers. 
We applied our approach to predict the output expectation values of quantum circuits describing the Trotter-decomposed dynamics of quantum Ising models, similarly to recent investigations on quantum utility experiments~\cite{ibm2023}.
We considered the connectivity allowed by the Guadalupe IBM chip, accounting for hardware errors via the \textit{FakeGuadalupe} noise model implemented in the Qiskit library.
%

In detail, the inputs of our CNNs include single-qubit noisy output expectation values, beyond the classical circuit descriptors -- in this study, rotation angles and qubit indices -- which were already considered in previous supervised learning studies.
Training and testing circuits are implemented across various regions of the physical chip. This strategic arrangement enables the CNN to visualize and learn from all potential connections between physical qubits during the training process.
Two circuit configurations were addressed, featuring either intra-layer or inter-layer random variations of the single-qubit rotation angles. 
The former angle configuration was already shown to be amenable to supervised learning~\cite{cantori2024challenges}. Yet, here we found that the inclusion of noisy expectation values leads to systematically superior performances. In the second configuration the boost is extreme. Indeed, while supervised learning with only classical descriptors drastically fails, the combination with noisy quantum circuit outputs leads to accurate predictions. A modified error model was implemented to allow us tuning the noise, and we quantified how the synergetic predictions improve when the quantum expectation values become more precise.

Notably, the CNNs trained (also) on noisy expectation values produce results more efficiently and with greater accuracy than a prominent error mitigation method, namely, ZNE. 
%
%
Moreover, our approach is a viable alternative to the one presented in Ref.~\cite{ibmML}, which relies on noisy expectation values mitigated via ZNE as training target values. Transfer-learning from small-scale to large scale circuits is a key feature of our network, allowing the prediction of expectation values for larger circuits than those in the training set, without the requirement for target values at these larger sizes.
In general, our strategy enables the integration of the strengths of classical deep learning and of noisy quantum computers, potentially outperforming  existing quantum error mitigation methods.



\begin{appendices}

\section{Manipulation of the noise strength associated with the CNOT gates}\label{appendix_noise}
\begin{figure*}
	\centering
	\includegraphics[width=0.5\textwidth]{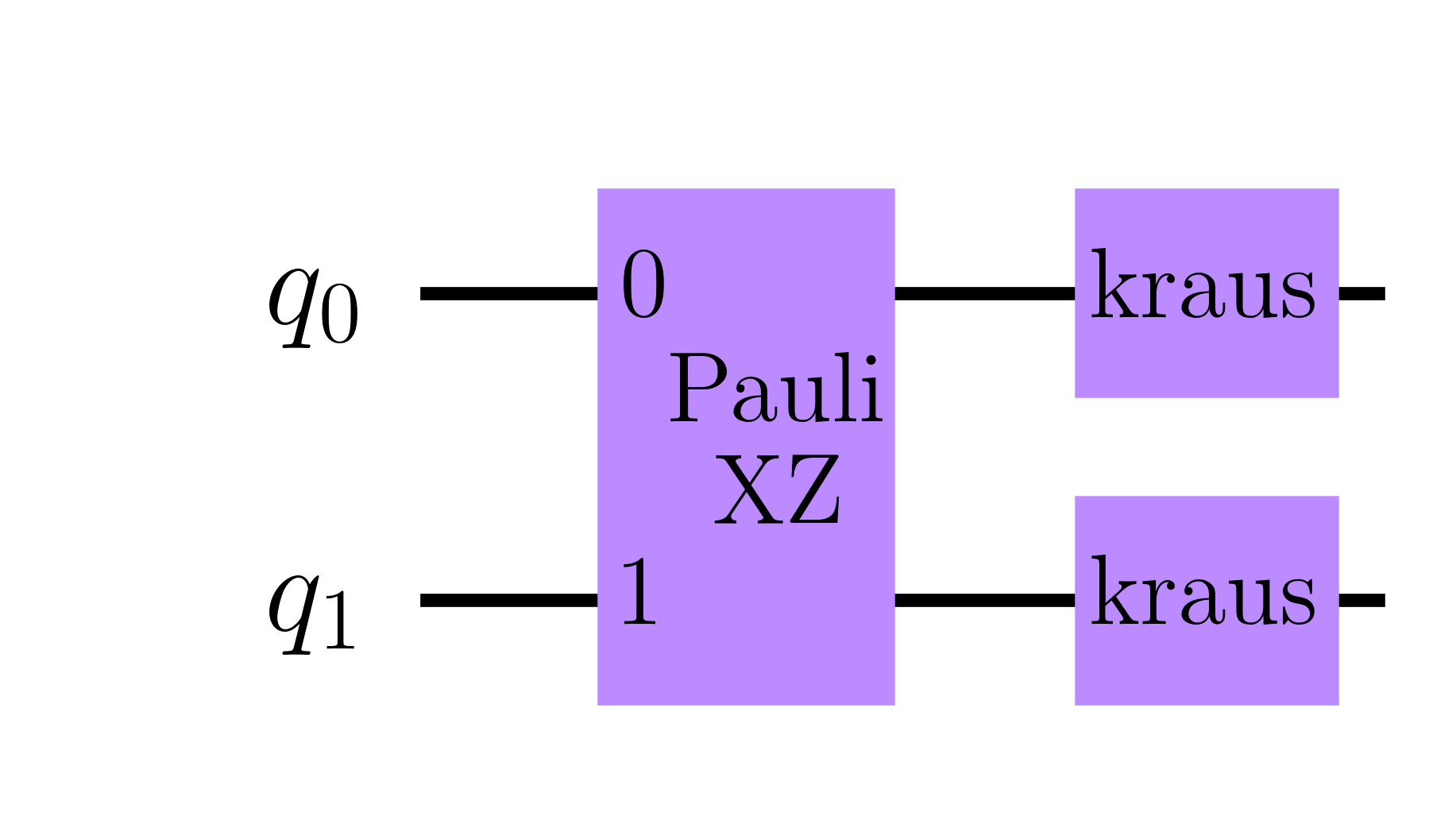}
	\caption{Noise associated with a CNOT gate in the \textit{FakeGuadalupe} noise model. The operators shown in the figure act on the qubit pair formed by the physical qubit 12 and the physical qubit 15 of the quantum chip (see Fig.~\ref{Fig1}). The probability that these operators act after a CNOT gate applied between these two qubit is $2.7\times10^{-4}$. The first operator applied to both qubits is the tensor product between the Pauli operators $X$ and $Z$, i.e. $X\otimes Z$. Then, different Kraus maps are applied to both qubits.
 }
	\label{error_qc}
\end{figure*}

In the noise model of \textit{FakeGuadalupe}, the CNOT noise is represented by a set of operators applied after each CNOT gate with varying probabilities. An example is depicted in Fig.~\ref{error_qc}. The specific operators and their associated probabilities depend on the considered pair of qubits. To tune the noise strength and the corresponding error, we introduce a circuit containing one identity gate per qubit into each circuit set, assigning it a probability of $1-p_{\mathrm{noise}}$ to take effect. Therefore, the remaining circuits, representing CNOT errors, collectively carry a probability of $p_{\mathrm{noise}}$ to occur after the CNOT application.
\section{Zero-noise extrapolation}
\label{appendixZNE}
To apply ZNE to the noisy expectation values, we utilize the Mitiq python library~\cite{mitiq}. Specifically, we employ Richardson extrapolation with noise scale factors  $\lambda=1,2,3$. To manipulate the noise level, the unitary folding map $G \rightarrow GG^\dagger G$ is applied to all the gates of the investigated quantum circuits for $\lambda=3$, and to a half of the gates (randomly selected) for $\lambda=2$.

\end{appendices}

\bmhead{Acknowledgements}
This work was supported by the PNRR MUR Project No. PE0000023-NQSTI and by the Italian Ministry of University and Research under the PRIN2022 project ``Hybrid algorithms for quantum simulators'' No. 2022H77XB7.
S.P. acknowledges support from the CINECA award IsCb2\_NEMCASRA, for the availability of high-performance computing resources and support.
S.P. also acknowledges the EuroHPC Joint Undertaking for awarding access to the EuroHPC supercomputer LUMI, hosted by CSC (Finland) and the LUMI consortium through a EuroHPC Benchmark Access call.

\bibliography{mybibliography}

\end{document}